\documentclass[pra,twocolumn,showpacs,groupedaddress,superscriptaddress,floatfix,aps,10pt]{revtex4-2}
\usepackage{bm,graphicx,amsmath}
\usepackage{placeins}
\usepackage{amssymb}
\usepackage{amsmath}
\usepackage{epsfig}
\usepackage{nicefrac}
\usepackage{multirow}
\usepackage{mathtools}
\usepackage{color}
\usepackage[colorlinks,linkcolor=blue,citecolor=blue,urlcolor=blue]{hyperref}
\usepackage{multirow}
\usepackage{siunitx}
\usepackage[utf8]{inputenc}
\usepackage{pgfplots}
\DeclareUnicodeCharacter{2212}{−}
\usepgfplotslibrary{groupplots,dateplot}
\usetikzlibrary{patterns,shapes.arrows}
\pgfplotsset{compat=newest}

\begin{document}
%
\title{Trapping of Bose-Einstein condensates in a three-dimensional dark focus\\
generated by conical refraction}
\author{D. Pfeiffer}
\author{L. Lind}
\author{J. K\"uber}
\author{F. Schmaltz}
\affiliation{Technische Universit\"at Darmstadt, Institut f\"ur Angewandte Physik, Schlossgartenstra{\ss}e 7, D-64289 Darmstadt, Germany}
\author{A. Turpin}
\affiliation{Departament de F\'isica, Universitat Aut\`onoma de Barcelona, E-080193 Bellaterra, Spain}
\affiliation{iLoF - Intelligent Lab on Fiber, Oxford OX1 2EW (UK)}
\author{V. Ahufinger}
\author{J. Mompart}
\affiliation{Departament de F\'isica, Universitat Aut\`onoma de Barcelona, E-080193 Bellaterra, Spain}
\author{G. Birkl}
\email[]{Contact for correspondence: apqpub@physik.tu-darmstadt.de}
\homepage[]{https://www.iap.tu-darmstadt.de/apq}
\affiliation{Technische Universit\"at Darmstadt, Institut f\"ur Angewandte Physik, Schlossgartenstra{\ss}e 7, D-64289 Darmstadt, Germany}
\affiliation{Helmholtz Forschungsakademie Hessen f\"ur FAIR (HFHF), Campus Darmstadt, Schlossgartenstra\ss e 2, 64289 Darmstadt, Germany}
\date{\today}
\begin{abstract}
We present an efficient three-dimensional dark-focus optical trapping potential for neutral atoms and Bose-Einstein condensates. This \lq\lq optical bottle\rq\rq\ is created by a single  blue-detuned light field exploiting the phenomenon of conical refraction occurring in biaxial crystals. The conversion of a Gaussian input beam to the bottle beam has an efficiency of close to 100\% and the optical setup requires the addition of the biaxial crystal and a circular polarizer only. Based on the conical-refraction theory, we derive the general form of the potential, the trapping frequencies, and the potential barrier heights. We present experiments on confining a $^{87}$Rb Bose-Einstein condensate in three dimensions. We determine the trap shape, the vibrational frequencies along the weak axis, as well as the lifetime of ultracold atoms in this type of potential.\\

\noindent
\textcolor{blue}{
Published as: Phys. Rev. A \textbf{108}, 053320 (2023), \url{https://doi.org/10.1103/PhysRevA.108.053320}
}
\end{abstract}


\maketitle
\section{Introduction}
\begin{figure}[hb!]
\centering
\includegraphics[width=1 \columnwidth]{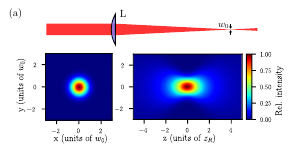}
\includegraphics[width=1 \columnwidth]{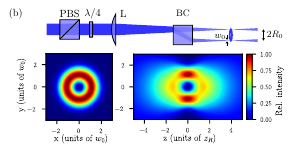}
\caption{(a) Schematic setup for a red-detuned attractive potential for neutral atoms provided by a focused Gaussian beam and density plots of relative intensity for the focal plane $(x,y,z=0)$ and along the propagation of the beam $(x=0,y,z)$. 
(b) Schematic setup for a blue-detuned repulsive potential provided by the conical-refraction bottle beam and the respective density plots of relative intensity}
\label{fig2}
\end{figure}

The position-dependent energy shift in an inhomogeneous light field can be used to efficiently trap ultracold atoms and Bose-Einstein condensates (BECs) in an optical dipole trap
\cite{nobel:opticaltrapping,grimm:1999:dipole,steck:2007:quantum}.
Atoms are attracted to regions of high intensity when using light whose frequency $\omega_\text{L}$ is lower than the relevant atomic transition frequency $\omega_0$ (red-detuned case with detuning $\Delta \coloneqq \omega_\text{L}-\omega_0<0$).
These attractive optical potentials are the most widely used conservative optical traps due to their simplicity since only a single tightly focused laser beam is needed, producing a strong intensity variation as depicted in Fig.~\ref{fig2}(a). On the other hand, red-detuned dipole potentials introduce enhanced light scattering onto the trapped atoms which are localized at the intensity maximum. This causes heating of the atoms, loss of particles from the BEC quantum state, and reduces the fidelity of quantum operations and high-precision measurements due to loss of coherence.\\
%
In contrast, blue-detuned optical traps ($\Delta>0$), allowing for a confinement of atoms at a local intensity minimum, have substantively decreased scattering rates and decoherence for atoms cooled close to the minimum of the trap potential. Therefore, they are ideal traps for high-precision experiments, but in most realizations demand more complex light-field geometries. Blue-detuned optical potentials have been used, e.g., for atomic clocks \cite{takamoto:2009:prl}, quantum information processing with neutral atoms in Rydberg states \cite{zhang:2011:pra,PhysRevA.88.013420,barredo:2020:rydberg}, and the investigation of Bose-Einstein condensates \cite{gaunt:2013:prl,turpin:2015:}.
Ideally, the local minimum where atoms are trapped has null intensity and is surrounded by steep intensity gradients which create the confining potential. Such light beams are referred to as \lq\lq optical bottle beams\rq\rq\ \cite{padgett:2000:ol}. Different methods have been reported for generating optical bottle beams, such as scanning blue-detuned laser beams for time-averaged potentials \cite{rudy:2001:oe}, conical lenses \cite{axicon:bottle:2001}, interfering Laguerre-Gaussian beams \cite{Puppe:2007:PhysRevLett.99.013002,isenhower:2009:ol}, phase holograms generated by phase plates \cite{Ozeri:99:PhysRevA.59.R1750} or spatial light modulators (SLMs) \cite{Xu:10}, crossing two or more vortex beams \cite{isenhower:2012:ol}, or using optical c-cut uniaxial and biaxial crystals \cite{krolikowski:2013:josab}. However, most of these methods have associated limitations such as the requirement of an extremely precise control of the optical elements needed to generate and align the complex beam geometry, limited conversion efficiency, or the fact that the intensity minimum is not exactly equal to zero \cite{zhan:2009:aop}.\\
In previous work, some of us have reported on the generation of a bottle beam with a point of exact null intensity, i.e., with a three-dimensional (3D) dark focus by using a biaxial crystal (BC) exploiting the phenomenon of conical refraction (CR) \cite{loiko:2013:ol}.
This configuration, shown in Fig.~\ref{fig2}(b), is almost as simple as a focused red-detuned Gaussian beam since it only requires the addition of a BC and potentially of a linear polarizer (polarizing beam splitter, PBS) and a quarter-wave plate ($\lambda / 4$) for creating circularly polarized light.
Focusing this beam through the biaxial crystal transforms a single focal spot into a ring-shaped intensity distribution in the focal plane. The ring radius $R_\mathrm{0}$ is given by the properties of the crystal and the radial extent of the intensity distribution is related to the focal spot size $w_\mathrm{0}$.
By choosing the appropriate ratio between $R_\mathrm{0}$ and $w_\mathrm{0}$, the emerging light field can be adjusted to the bottle beam configuration.
First results on trapping of ultracold atoms \cite{kuber2014dynamics} and confinement of absorbing droplets based on such a dark focus have been shown by Esseling \textit{et al}. \cite{esseling:2018:bb}.
%
In the present article, we report the trapping of ultracold atoms and BECs in such a 3D dark-focus beam and
analyze the implementation in detail both experimentally and theoretically.
%
In Sec. \ref{sec2}, we present the main characteristics of the CR phenomenon, its theoretical basis, and the properties of the 3D dark-focus beam. In Sec. \ref{sec3}, we apply the harmonic approximation around the dark focus and derive expressions for trapping frequencies and heights of the potential barriers as a function of the parameters of the trap configuration. We characterize the experimental intensity distribution of the bottle beam and determine the properties of the trapping potential in Sec. \ref{sec4}. 
In Sec. \ref{sec5}, we demonstrate the trapping of a BEC of $^{87}$Rb, and discuss different configurations for further applications in Sec. \ref{sec:evaluation}. We sum up the main conclusions of this work in Sec. \ref{conclusions}.
\section{Conical refraction}
\label{sec2}
The CR phenomenon (for a review see Ref. \cite{Turpin:review:2016}) transforms an unpolarized or circularly polarized input light beam focused to a waist $w_0$ into a bright ring of radius $R_0$ at the focal plane when the input light beam passes along one of the optical axes of a biaxial crystal \cite{belski:1978:os,berry:2006:prsa,berry:2007:po,kalkandjiev:2008:spie}, as can be seen in  Fig.~\ref{fig2}(b). The CR ring radius $R_0= l \tan \alpha \simeq l \alpha$ is the product of the crystal length $l$ and the CR semi angle $\alpha$ \cite{kalkandjiev:2008:spie}. The CR semi angle $\alpha$ depends on the three principal refractive indices of the crystal as $\alpha \simeq \sqrt{(n_2-n_1)(n_3-n_2)}/n_2$, where it is assumed that $n_{1}<n_{2}<n_{3}$. At any point of the CR ring, the electric field is linearly polarized with the polarization axis rotating so that every pair of diagonally opposite points have orthogonal polarization. The absolute orientation of the polarization distribution depends on the orientation of the plane of the optical axes of the crystal \cite{kalkandjiev:2008:spie,turpin_EBs:2013:oe}.\\  
The theoretical model describing the beam propagation in CR is based on the Belsky--Khapalyuk--Berry (BKB) integrals \cite{belski:1978:os,berry:2007:po}. 
For an input light beam with electric field $\mathbf{E}_{\text{in}}$, wavenumber $\kappa=\frac{2\pi}{\lambda}$, and using the cylindrically symmetric two-dimensional (2D) Fourier transform $a\left( \kappa \right)=2\pi\int_{0}^{\infty} r E_{\rm{in}} \left( r \right) J_{0} \left( \kappa r \right) \mathrm{d} r$, the normalized BKB integrals in cylindrical coordinates can be written as \cite{turpin:2015:} 
\begin{align}
B_{\rm{C}}(\rho,Z) =&\frac{1}{2\pi}\int^{\infty }_{0} \eta a \left( \eta \right) \mathrm{e}^{-\mathrm{i} \frac{Z}{4} \eta ^{2} } \nonumber \\ 
& \times \cos \left( \eta \rho_{0}\right) J_{0}\left( \eta \rho \right) \mathrm{d}\eta  \label{Bc} \\
B_{\rm{S}}(\rho,Z)=& \frac{1}{2\pi}\int^{\infty }_{0} \eta a \left( \eta \right) \mathrm{e}^{-\mathrm{i} \frac{Z}{4} \eta ^{2} } \nonumber \\ 
& \times\sin \left( \eta \rho_{0}\right) J_{1}\left( \eta \rho \right) \mathrm{d} \eta \label{Bs}
\end{align}
using the normalized coordinates given in Table~\ref{tab:coordinates}.
\begin{table}[h]
    \centering
    \begin{tabular}{cc}
    \hline\hline
    \rule{0pt}{12pt}
        $\rho = \frac{r}{w_0}$ & $X = \frac{x}{w_0}$ \\
    \rule{0pt}{12pt} 
        $\rho_0 = \frac{R_0}{w_0}$ & $Y = \frac{y}{w_0}$ \\
    \rule{0pt}{12pt} 
        $z_{\mathrm{R}} = \frac{\pi w_0^2}{\lambda}$ & $Z = \frac{z}{z_{\rm R}}$
    \vspace{3pt}\\
    \hline\hline
    \end{tabular}
    \caption{Normalized coordinates of the theoretical model describing conical refraction. Here, $z_{\text{R}}$ denotes the Rayleigh range of the incoming Gaussian beam propagating along the positive $z$ axis. The control parameter $\rho_0$ serves as a characteristic number for describing the structure of the appearing intensity distribution.}
    \label{tab:coordinates}
\end{table}
Here, $\eta=\kappa w_0$ and $J_{\alpha}$ is the $\alpha^{\text{\rm th}}$-order Bessel function of the first type.
The CR intensity distribution for a circularly polarized input beam is given by
\begin{equation}
I(\rho,Z) = \left| B_{\rm{C}}(\rho,Z) \right|^{2} + \left| B_{\rm{S}}(\rho,Z) \right|^{2}.\quad \label{Eqs_output_beam_intensity_CP}
\end{equation}
This can be derived from a Gaussian input beam with power $P$ and normalized transverse profile of the electric field amplitude $E(\rho)=\sqrt{2P/\pi w_0^{2}} \exp (-\rho^2)$ using the 2D Fourier transform $ a \left( \eta \right)=\sqrt{2P\pi /w_0^2}\exp (-\eta^{2}/ 4)$. 
Depending on the parameter $\rho_0$, the shape of the annular CR intensity pattern can vary drastically. Therefore, $\rho_0$ will be used as a control parameter. 
For $\rho_0 \ll 1$, the light field results in a single focused spot. 
For $\rho_0 \gg 1$, the intensity pattern at the focal plane will form two bright rings separated by a dark (Poggendorff) ring
\cite{phelan:2009:oe,peet:2010:oc,turpin:2012:ol,turpin_rings:2013:ol,peinado:2013:ol,turpin_vault:2013:oe}.
This pattern of concentric rings, while illuminated with blue-detuned light, can be used to create a repulsive ring-shaped guiding potential for cold atoms and BECs \cite{turpin:2015:}. Finally, for intermediate values of $\rho_0 \approx 1$, the structure of the CR beam changes substantially, giving rise to different optical ring potentials including a dark focus \cite{loiko:2013:ol,peet:2010:oe,turpin:2014:ol}. 
\subsection*{Generation of a 3D dark focus using CR}
\label{sec2a}
For the value $\rho_0 = \rho^{\rm{DF}}_0 = 0.924$, it has been shown that the CR beam possesses a 3D dark focus of vanishing intensity at the focal plane and increasing intensity in all directions \cite{padgett:2000:ol,loiko:2013:ol}. Thus, it is a perfect bottle beam  as shown in Fig.~\ref{fig2}(b).
In this case, the transverse profile in the focal plane is formed by a doughnut-like intensity distribution [Fig.~\ref{fig2}(b), lower left].
Along the axis of propagation, the on-axis intensity at first increases with distance from the focal plane and drops again for larger distances, as shown in Fig.~\ref{fig2}(b), lower right. 
At the focal plane, the positions of maximum intensity form a ring with radius $\rho_{\rm{max}} = 1.096$ and peak intensity $I(\rho = 1.096, Z=0) = 0.199\times I_0$ with $I_0=\frac{2P}{\pi w_0^2}$ being the peak intensity of the Gaussian input beam in the focal plane without the CR crystal. 
Along the axial direction, there are two points of maximum intensity at $Z_{\rm{max}} = \pm 1.388$ with intensity $I(\rho = 0, Z=\pm 1.388) = 0.138\times I_0$. The intensity reaches 50\% of the maximum value at position $Z=\pm 0.598$ along the axis, and the minimum in the radial direction vanishes at $Z=\pm 1.174$. 
The weakest point of the bottle, thus defining the trap depth, is located at position $\rho=0.619, Z= \pm1.11$ with a value of $I(\rho = 0.619, Z=\pm 1.11) = 0.133\times I_0$.
Therefore, under the condition $\rho_0 = \rho^{\rm{DF}}_0 = 0.924$, a circularly polarized input beam is transformed into a perfect optical bottle beam with a point of exact null intensity at $\rho, Z = 0$ equal to $x=y=z=0$. 
This feature makes this CR beam an ideal candidate for atom trapping experiments with blue-detuned light, in simplicity corresponding to a red-detuned trap generated by a focused Gaussian beam. The values for the intensity levels at specific positions are summarized in Table~\ref{tab:intensities}.
\begin{table}[h!]
    \centering
    \begin{tabular}{r|ll}
    \hline\hline
    \multicolumn{1}{c}{Intensity} \vline& \multicolumn{2}{c}{Position} \\
    \hline
        $I_{\text{r,max}} = 0.199~I_0$ & $\rho=1.096$ &  $Z=0$ \\
        $I_{\text{z,max}} = 0.138~I_0$ & $\rho=0$ & $Z=\pm 1.388$\\
        $I_{\text{trap}} = 0.133~I_0$ & $\rho=0.619$ & $Z=\pm1.11$\\
        $I_{\text{fork}} = 0.134~I_0$ & $\rho=0$ & $Z=\pm1.174$\\
        $I_{\text{50\%,z}} = 0.069~I_0$ & $\rho=0$ & $Z=\pm0.598$\\
    \hline\hline
    \end{tabular}
    \caption{Specific intensity values and corresponding positions characterizing the dark-focus beam for $\rho_0 = \rho^{\rm{DF}}_0 $ in reference to the maximum intensity $I_0$ of a focused Gaussian beam.}
    \label{tab:intensities}
\end{table}
\section{Theoretical formulation of the properties of the 3D dark focus}
\label{sec3}
In this section, we study the behavior of the CR beam close to the trap center, i.e., for 
$\rho \approx 0$, $Z \approx 0$, and for the value $\rho_0 = \rho_0^{\rm{DF}} = 0.924$.
We deduce the height of the potential barriers and, using a harmonic approximation, the vibrational frequencies for trapping of ultracold atoms for the example of the alkali atom $^{87}$Rb. The strength of the dipole potential \cite{grimm:1999:dipole,steck:2007:quantum} will be considered for typical experimental conditions as
\begin{align}
U(\mathbf{r}) =& I(\textbf{r}) \tilde{U}_0 \label{potential} \\
\tilde{U}_0 =& \frac{\pi c^2}{2} \left[ \frac{\Gamma_{\rm D_2}}{\omega^3_{\rm D_2}} \left( \frac{2}{\omega_{\rm L}-\omega_{\rm D_2}} \right) \right. \nonumber \\
&+ \left. \frac{\Gamma_{\rm D_1}}{\omega^3_{\rm D_1}} \left( \frac{1}{\omega_{\rm L}-\omega_{\rm D_1}} \right ) \right].\label{auxpot} 
\end{align}
In $\tilde{U}_0$ we have applied the rotating-wave approximation. 
Here, $c$ is the speed of light in vacuum, $\Gamma_{\rm{D}_{\mathit i}}$ and $\omega_{\rm{D}_{\mathit i}}$ ($ i\in \left\{1,2\right\}$) are the natural line width and frequency of the $\rm{D}_{\mathit i}$ line of the atomic species, and $\omega_{\rm L}$ is the frequency of the input light. 
The spatial intensity distribution $I(\textbf{r})$ is given by Eq.~\eqref{Eqs_output_beam_intensity_CP}.
\subsection{Radial direction}
The Taylor series of the Bessel functions of order $\alpha$, $J_{\alpha}(\eta\rho)$, around $\eta\rho=0$ can be written as 
\begin{equation}
J_{\alpha}(\eta\rho) = \sum^{\infty}_{k=0} \frac{(-1)^{k}}{k! \Gamma(k+\alpha+1)} \left( \frac{\eta\rho}{2}\right)^{2k+\alpha},
\label{bessel}
\end{equation}
where $\Gamma(t) = \int^{\infty }_{0} x^{t-1} \mathrm{e}^{-x} dx$ is the Gamma function. Under this expansion and for an input beam with Gaussian profile, Eq.~\eqref{Bc} with $\alpha=0$ and Eq.~\eqref{Bs} with $\alpha=1$ can be rewritten as
\begin{align}
B_{\rm{C}}(\rho,Z) =& \sqrt{\frac{P}{2\pi w(Z)^2}} \int^{\infty}_{0} \eta \mathrm{e}^{\frac{-\eta^2 (1+\mathrm{i}Z)}{4}} \cos\left( \eta \rho_0 \right) \nonumber \\  
& \times \sum^{\infty}_{k=0} \frac{(-1)^{k}}{k! \Gamma(k+1)} \left( \frac{\eta \rho}{2}\right)^{2k} \mathrm{d} \eta, 
\label{Bc_bessel} \\
B_{\rm{S}}(\rho,Z)=& \sqrt{\frac{P}{2\pi w(Z)^2}} \int_{0}^{\infty} \eta \mathrm{e}^{\frac{-\eta^2 (1+\mathrm{i}Z)}{4}} \sin\left( \eta \rho_0 \right) \nonumber \\ 
& \times \sum^{\infty}_{k=0} \frac{(-1)^{k}}{k! \Gamma(k+2)} \left( \frac{\eta \rho}{2}\right)^{2k+1} \mathrm{d} \eta. \label{Bs_bessel}
\end{align}
Equations (\ref{Bc_bessel}) and (\ref{Bs_bessel}) can be analytically solved, obtaining the following expressions: 
\begin{align}
B_{\rm{C}}(\rho,Z)=& \sqrt{\frac{2 P}{\pi w(Z)^2}} \sum^{\infty}_{k=0} \frac{(-1)^{k} \rho^{2k}}{k! (1+\mathrm{i}Z)^{k+1}}  \nonumber \\
& \times {}_1F_1 \left( k+1;\frac{1}{2}; \frac{-\rho_0^2}{1+\mathrm{i}Z} \right), \label{Bc_final} \\
B_{\rm{S}}(\rho,Z)=& \sqrt{\frac{8 P \rho_0^2}{\pi w(Z)^2}} \sum^{\infty}_{k=0} \frac{(-1)^{k} \rho^{2k+1}}{k! (1+\mathrm{i}Z)^{k+1}} \nonumber \\ & \times   {}_1F_1 \left( k+2;\frac{3}{2}; \frac{-\rho_0^2}{1+\mathrm{i}Z} \right), \label{Bs_final} 
\end{align}
where ${}_{1}F_{1}(a;b;z)$ is the Kummer confluent hyper geometric function \cite{mathews:2021:physicist}. 
This formulation is valid for all values of $\rho_{0}$ for which the point of minimum intensity remains at $\rho=0$.
%
For the 3D dark-focus beam ($\rho_0 = 0.924$), the minimum intensity in the focal plane is zero. We find that for $\rho \lesssim 1$, the expression for $k=0$ is a good approximation to the full CR intensity profile. 
Therefore, it is sufficient to keep the $k=0$ terms of the series in Eqs.~\eqref{Bc_final} and \eqref{Bs_final} only. 
In this case, the intensity of the CR beam reads
\begin{align}
I \left(\rho \lesssim 1, Z\right) 
=& \frac{2P}{\pi w(Z)^2}  \left( \left| \frac{ _1F_1 \left( 1;\frac{1}{2}; \frac{- \rho_{0}^2}{1+\mathrm{i}Z} \right)}{1+\mathrm{i}Z} \right|^2 \right. \nonumber 
\\  &+ \left.4 \rho_0^2 \left| \rho^2 \frac{ _1F_1 \left( 2;\frac{3}{2}; \frac{- \rho_{0}^2}{1+\mathrm{i}Z} \right)}{(1+\mathrm{i}Z)^2} \right|^2 \right)
\label{intapproxr}
\end{align}
with $w(Z) = w_0\sqrt{1+Z^2}$.
The first term in Eq.~\eqref{intapproxr} is an offset to the potential that appears for $Z \neq 0$, as shown in Fig.~\ref{fig2}(b), lower right. As a consequence, trapping atoms outside of the focal plane increases spontaneous scattering. To obtain the trapping frequencies of the potential in the radial direction, one applies the harmonic approximation to the second term of Eq.~(\ref{intapproxr}), which yields
\begin{equation}
\omega_{r} (Z)= \sqrt{\frac{16 \rho_0^2 \tilde{U_0} P}{\pi m w_0^4 (1+Z^2)}}\left| \frac{ _1F_1 \left( 2;\frac{3}{2}; \frac{- \rho_0^2}{1+\mathrm{i}Z} \right)}{(1+\mathrm{i}Z)^2} \right|,
\label{wr}
\end{equation}
where $m$ is the atom mass. Note that this approximation is valid only in the region where the optical bottle is formed, i.e., for $Z \in [-1.174,1.174]$. In Eq.~(\ref{wr}), we have undone the normalization of the radial coordinate, i.e., we have replaced $\rho$ by $r/w_0$.
For the 3D dark focus, at the focal plane $(Z = 0)$ the radial trapping frequency reads
\begin{equation}
\omega_{r} (\rho_0 = 0.924, Z = 0)= 0.383 \times\sqrt{\frac{8 \tilde{U_0} P}{\pi m w_0^4}}
\label{wr_rho0}
\end{equation}
The potential barrier along the radial direction, i.e., at the position $\textbf{r} =(\rho=1.096, Z=0)$, is not well described by the harmonic approximation. To give a full description of the radial maximum, expressions up to at least $k=4$ must be considered 
in Eqs.~(\ref{Bc_final}) and (\ref{Bs_final}).
The value of the radial barrier is
\begin{align}
U_{\text{r,max}}=U(\rho=1.096, Z=0) &= 0.199 \times \tilde{U_0} \frac{2P}{ \pi w_0^2} \nonumber\\&= 0.199 \times \tilde{U_0}I_0
\label{Ur}
\end{align}
\subsection{Axial direction}
In the axial direction, a compact expression for any value of $\rho_0$ cannot be obtained.
For the case of $\rho_0=\rho^{\rm{DF}}_0$, the point of minimum intensity is at $\rho=0$. Here, the approximation from Eq.~(\ref{bessel}) used before is not needed
since $J_1(0) = 0$ and $J_0(0) = 1$ and, as a consequence, $B_{\rm{S}} \left( \rho=0, Z\right) = 0$. Therefore, the intensity is solely given by $B_{\rm{C}}$ as follows:
\begin{align}
I \left( \rho=0, Z\right) =& \left| B_{\rm{C}} \left( \rho=0, Z \right) \right|^2  \label{intapproxz} \\
=& \frac{P}{2\pi w_0^2} \left|\int^{\infty }_{0} \eta \mathrm{e}^{-\frac{\eta^{2}(1+\mathrm{i}Z)}{4}} \cos \left( \eta \rho_{0}\right) \mathrm{d}\eta \right|^2  \nonumber\\ 
=& \frac{2P}{\pi w_0^2 \left| 1+\mathrm{i}Z \right|^2} \left| 1 - \frac{2 \rho_0 D\left( \frac{\rho_0}{\sqrt{1+\mathrm{i}Z}}\right)}{\sqrt{(1+\mathrm{i}Z)}} \right|^2, \nonumber 
\end{align}
where $D(x)$ is the Dawson function \cite{mccabe:1974:continued}. The second order of the Taylor series of this analytical solution leads to the following expression for the trapping frequency $\omega_{z}$ along the axial direction:
\begin{equation}
\omega_{z}(\rho_0=0.924, \rho=0) = \sqrt{\frac{\tilde{U_0} P}{\pi m w_0^2 z_{\rm{R}}^2}} = \sqrt{\frac{\tilde{U_0} P\lambda^2}{\pi^3 m w_0^6}}.
\label{wz}
\end{equation}
The height of the potential barriers along the axial direction, i.e., at $\textbf{r} =(\rho=0, Z=\pm 1.388)$, can be directly obtained from Eq.~(\ref{intapproxz}) as
\begin{align}
U_{\text{z,max}}=U(\rho=0, Z=\pm 1.388) &= 0.138 \times \tilde{U_0} \frac{2P}{\pi w_0^2} \nonumber\\ &= 0.138 \times \tilde{U_0}I_0.
\label{Uz}
\end{align}
\section{Experimental realization of the dark focus}
\label{sec4}
\begin{figure}[t!]
\centering
\includegraphics[width=1 \columnwidth]{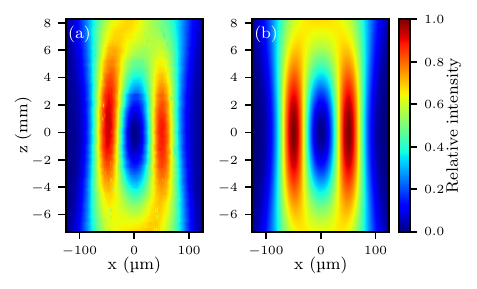}
\caption{Intensity distribution of the dark-focus bottle beam: (a) Two-dimensional density plot of the experimental intensity distribution in the $xz$ plane. 
(b) Corresponding calculated intensity distribution.} 
\label{Int_XZ_Plane}
\end{figure}

\begin{figure}[t!]
\centering
\includegraphics[width=1 \columnwidth]{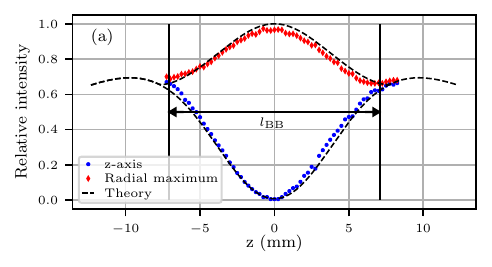}
\includegraphics[width=1 \columnwidth]{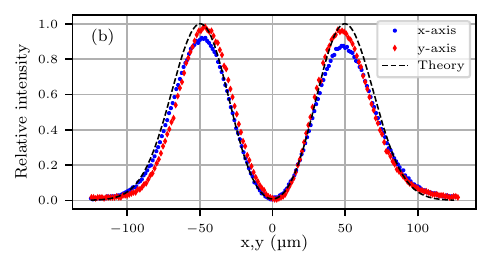}
\caption{(a) Maximum (red) and minimum (blue) values of the relative intensity of the CR pattern at different planes along the beam propagation axis $z$. 
Calculated values are shown as dashed lines. 
(b) Relative intensity along the $x$ and $y$ axis in the focal plane at $z=0$. 
The dashed line shows the calculated values.}
\label{fig_bottle_exp}
\end{figure}

In order to generate the experimental intensity distribution for a dark-focus bottle beam, in our setup a collimated Gaussian beam with a waist of $w_{\text{in}}=\SI{1150\pm 50}{\micro\meter}$ is used as input. 
The beam is focused with an achromatic lens of focal length $f_{\text{L}1}=\SI{200}{mm}$. 
A series of beam profiles at different positions along the beam axis are taken to determine spot size and Rayleigh range of the focused beam. 
The beam waist in the focal plane and the Rayleigh range are measured to $w_0=\SI{43.6\pm 3}{\micro\meter}$ and $z_{\mathrm{R}} = \SI{6.85\pm 0.6}{mm}$, respectively. 
When a quarter-wave plate and a potassium gadolinium tungstate [KGd(WO$_4)_2$] crystal are placed in the beam path, the expected dark focus emerges [see Fig.~\ref{Int_XZ_Plane}(a)]. 
Given the length $l=\SI{2.2 \pm 0.1}{mm}$ and semi angle $\alpha=\SI{19.07}{mrad}$ of the crystal, $R_0=\SI{42.0\pm 1.9}{\micro\meter}$ and the CR parameter $\rho_0=\SI{0.96\pm 0.08}{}$, which is close to the targeted $\rho_{0}^{\rm DF}=0.924$. 
For characterization of the intensity distribution of the dark focus, a series of 63 images of the $xy$ plane are taken along the beam propagation axis with a mutual separation of $\SI{250\pm 10}{\micro\meter}$. 
In Fig.~\ref{Int_XZ_Plane}(a), cuts through the center of the images along the $x$ axis are stacked along the image position $z$, displaying the intensity distribution in the $xz$ plane.
The density plot shows the dark focus generated by CR. 
For comparison, we calculate the corresponding intensity distribution based on Eqs.~\eqref{Bc}--\eqref{Eqs_output_beam_intensity_CP} for the measured beam parameters. 
The resulting density plot is shown in  Fig.~\ref{Int_XZ_Plane}(b).
The spatial distribution and the intensity values show good agreement. 
In the experimental plot, some deformations are visible which are due to imperfections in the incident beam and in the crystal and its alignment.\\
The minimal and maximal intensity as a function of $z$ are extracted from Fig.~\ref{Int_XZ_Plane}(a) and compared to the calculated values. 
We give the intensity along the $z$ axis for the center ($x=y=0$) of the trap in blue in Fig.~\ref{fig_bottle_exp}(a).
For the intensity maxima of each plane along $z$, we determine the mean value of the maximal intensity along the $x$ and $y$ axis, 
\begin{align}
I_{\max}\left(z\right) = \frac{\max\left[I\left(x,y=0,z\right)\right]+\max\left[I\left(x=0,y,z\right)\right]}{2}.
\end{align}
%
The resulting values are shown in red in Fig.~\ref{fig_bottle_exp}(a). 
Dashed lines represent the corresponding values calculated with Eqs.~\eqref{Bc}--\eqref{Eqs_output_beam_intensity_CP}. 
The experimental values show only small deviations from the calculated values, proving good agreement between theory and experiment.
From the intensity profiles in Fig.~\ref{fig_bottle_exp}(a) we extract the maximum length of the \lq\lq bottle\rq\rq\ $l_{\rm BB}$ where atoms can be confined. 
We define this as the separation of the two positions along $z$ where the potential minimum in the radial direction vanishes, allowing for a 5\% uncertainty in recognizing the disappearance of trapped atoms, i.e., 
$I(x=0,y=0,z)\geq 0.95 \times I_\text{max}(z)$ for $|z|\geq \frac{1}{2} l_{\rm BB}$. From Fig.~\ref{fig_bottle_exp}(a) we determine the length of the "bottle" as $l_{\rm BB}=\SI{14.2\pm 0,5}{mm}$.\\
We evaluate the radial confinement by comparing the experimental values of the intensities in the focal plane $z=0$ along the $x$ and $y$ axis to calculations. 
In Fig.~\ref{fig_bottle_exp}(b), the intensity values extracted from the experimental image along the $x$ (blue) and $y$ axis (red) are shown together with the calculated values (dashed). 
The measured 2D intensity distribution can be seen in the inset of Fig.~\ref{fig:setup}.
From the position of the radial intensity maxima $I_{\text{r,max}}$, in principle, the nominal value of the ring radius $R_0$ can be determined. Since the underlying numerical relation is a function of $\rho_0$, prior knowledge of $\rho_0$ is required which adds significant uncertainties and turns into circular reasoning. For $\rho_0=\SI{0.96\pm 0.08}{}$ we extract a value of $R_0=\SI{41.2\pm 2.8}{\micro\meter}$ from the experimental data of Fig.~\ref{fig_bottle_exp}(b). Due to the stated issues, we use the calculated value $R_0=\SI{42.0\pm 1.9}{\micro\meter}$ for the remainder of this paper.\\
For evaluating the feasibility of trapping atoms in a bottle beam based on the given intensity distribution, we calculate the trapping frequencies
in the focal plane after the crystal. It is important to notice that the CR dark focus can be reimaged to additional focal planes with preserved structure and scaled parameters. We give the scaling at the end of this section. 
Experimentally, an optical power of $\SI{52\pm 5}{mW}$ at $\SI{793.96}{\nano\meter}$ wavelength was available to generate the bottle beam. 
Taking the measured beam parameters ($\omega_0$, $z_{\mathrm{R}}$), the calculated ring radius $R_0$, and Eqs.~\eqref{wr_rho0} and \eqref{wz}, we get 
$\omega_{r}^{\rm th} = 2\pi\times\SI{295\pm 65}{\hertz}$ and 
{$\omega_{z}^{\rm th}=2\pi\times\SI{1.87\pm 0.23}{\hertz}$} in the focal plane of the CR for $^{87}$Rb atoms.
By performing harmonic fits to the axial intensity values in Fig.~\ref{fig_bottle_exp}(a) and to the radial values in the focal plane in Fig.~\ref{fig_bottle_exp}(b), we determine the mean curvature of the intensity distributions along both axes. Expressed as trapping frequencies for $^{87}$Rb atoms, we get $\omega_{r} = 2\pi\times\SI{298\pm10}{\hertz}$ and $\omega_{z} = 2\pi\times\SI{1,89\pm 0,11}{\hertz}$, in excellent agreement with the calculated values.\\
Since the focal plane after the CR crystal is outside the vacuum chamber (see Fig.~\ref{fig:setup}),
the dark-focus intensity distribution is relayed into the chamber with magnification $|M_{\rm I}|<1$.
This reimaging preserves the structure of the light field while scaling the dimensions according to the magnification $|M_{\rm I}|$. 
The beam waist and the ring radius scale with $|M_{\rm I}|$, the Rayleigh range and $l_{\rm BB}$ with $|M_{\rm I}|^2$, the radial trapping frequency $\omega_r$ with $1/|M_{\rm I}|^2$, and the axial trapping frequency $\omega_z$ with $1/|M_{\rm I}|^3$.\\
\section{3D Confinement of a rubidium BEC in the dark focus}
\label{sec5}
\subsection{Loading the dark-focus trap with atoms}
\begin{figure}[t!] 
\centering
\includegraphics[width=1 \columnwidth]{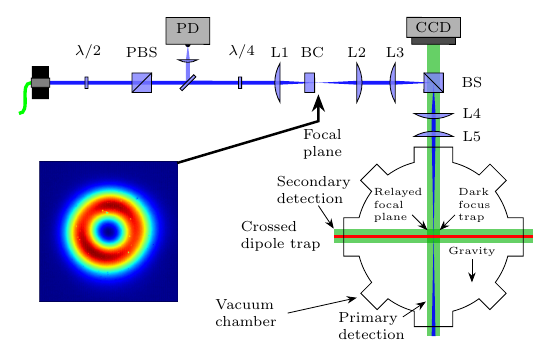}
\caption{Experimental setup for the creation of a dark-focus potential based on conical refraction. The axis of the bottle beam is oriented along the direction of gravity. The inset shows the transverse intensity distributions in the focal plane. 
Two detection paths can be used to image the atom distribution via absorption imaging. The primary path is aligned with gravity. The secondary path is in the plane of the crossed dipole trap oriented perpendicular to gravity. (BS: beam splitter; PBS: polarizing beam splitter; PD: photodiode; CCD: detection camera; BC: biaxial crystal; L1--L5: achromatic lenses)}
\label{fig:setup}
\end{figure}

In order to demonstrate the experimental ability of the CR dark focus to confine atoms in three dimensions, we transfer a BEC of $^{87}$Rb into the rescaled bottle-beam potential and measure the axial trapping frequency, the axial and radial extent of the potential minimum, and the lifetime of the cold atoms in the trap. 
A schematic of the experimental setup is shown in Fig.~\ref{fig:setup} together with the transverse intensity distribution in the focal plane. 
Two pairs of achromatic lenses reimage and rescale the intensity distribution emerging in the focal plane into the vacuum chamber. 
The $^{87}$Rb BEC of typically 25000 atoms with a temperature of $\SI{25}{nK}$ is produced in a crossed dipole trap (CDT) given by two intersecting focused laser beams of a $\SI{1070}{nm}$ fiber laser \cite{lauber:2011:pra}. 
The plane generated by the two beams of the CDT is normal to gravity and the propagation axis of the bottle beam.
A light sheet potential can be used in the plane of the CDT to provide an additional potential against gravity or to select a specific $z$ plane for analysis.
For the bottle-beam potential the light of a titanium:sapphire (TiSa) laser at a wavelength of $\lambda=\SI{793.96}{nm}$ is used.
The beam is intensity stabilized and a maximal optical power of $P=\SI{52\pm 5}{mW}$ in the vacuum chamber can be reached assuming no losses at the  AR-coated vacuum window. 
The light of the TiSa laser is delivered to the experiment by an optical fiber providing a collimated beam of waist $w_\text{in}$. 
To ensure a circular polarization of the incident beam, a combination of a polarizing beam splitter and a quarter-wave plate are used. 
The emerging intensity distribution in the focal plane (inset in Fig.~\ref{fig:setup}) is reimaged with a first telescope with $f_{\text{L}2}=\SI{400}{mm}$  and $f_{\text{L}3}=\SI{200}{mm}$ and finally mapped into the vacuum chamber with a second telescope with $f_{\text{L}4}=\SI{400}{mm}$  and $f_{\text{L}5}=\SI{300}{mm}$.
With the given geometry of the setup, the expected magnifications of the telescopes are $|M^{\mathrm{th}}_{2,3}|=\SI{0.5 \pm 0.01}{}$ and $|M^{\mathrm{th}}_{4,5}|=\SI{0.60 \pm 0.03}{}$ resulting in a total magnification of $|M^{\mathrm{th}}_{\rm tot}|=\SI{0.3 \pm 0.02}{}$.\\
To demonstrate the 3D trapping of atoms in the bottle beam, the dark focus is overlapped with the CDT. 
After the BEC is created in the CDT, the optical power of the bottle beam is increased linearly in three distinct temporal segments.
The power in the CDT is elevated during this time to ensure a sufficient potential against gravity. 
Initially, the power of the bottle beam is increased linearly up to $\SI{2.4}{mW}$ in $\SI{20}{ms}$ to adiabatically change the radial trapping potential. 
Only low optical power is needed to confine the atoms radially since the trap curvature  of the bottle beam  in this dimension is significantly larger than along the $z$ axis.
After this initial segment, the power of the bottle beam is linearly increased to $\SI{19}{mW}$ and further to $\SI{52}{mW}$ in two separate segments of $\SI{2}{ms}$ duration. 
After the final power is reached, the CDT is switched off and the atoms are trapped by the bottle beam alone. 
To maximize the transfer efficiency from the CDT to the dark-focus trap, it is necessary to align the minima of the potentials in the $x$ and $y$ direction. This prevents losses induced by scattering of photons resulting from the high radial intensity and small detuning from the D1 transition of the bottle beam.
\subsection{Characterization of the parameters of the dark-focus potential}
\begin{figure}[t!]
\centering
\includegraphics[width=1 \columnwidth]{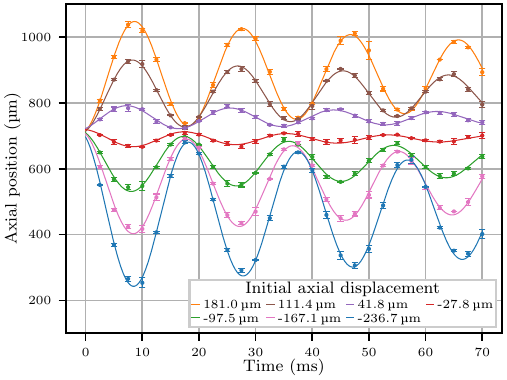}
\caption{Axial oscillations of the center of the atom distribution in the bottle-beam potential modified by gravity. Oscillations are recorded for varying initial displacements of the atoms in the CDT relative to the equilibrium position in the modified potential. Every data set (colored dots) is fitted with a damped sinusoidal function (colored lines) to extract the oscillation frequency.}
\label{fig:oscillations}
\end{figure}

\begin{figure}[h!]
\centering
\includegraphics[width=1 \columnwidth]{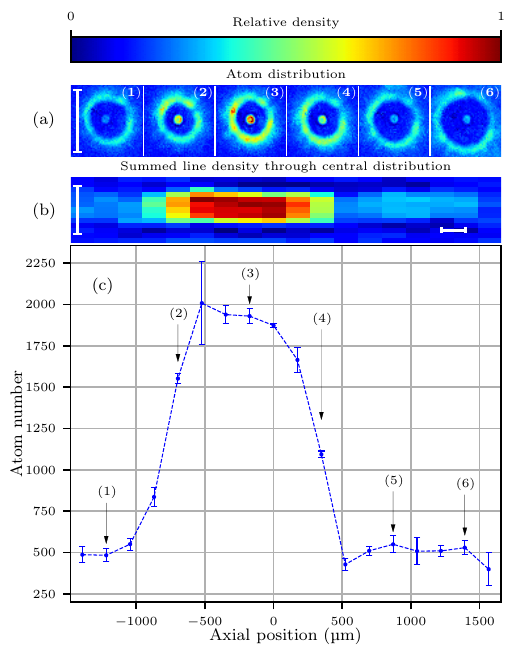}
\caption{(a) Density plots for six different planes along $z$. The white scale bar at left has a length of $\SI{250}{\micro\meter}$ for reference. The selected planes correspond to the positions $\SI{-1218}{\micro\meter}$, $\SI{-696}{\micro\meter}$, $\SI{-174}{\micro\meter}$, $\SI{348}{\micro\meter}$, $\SI{870}{\micro\meter}$, and $\SI{1392}{\micro\meter}$. Atoms that are not captured in the bottle beam are pushed outside by the blue-detuned light and can be seen as an annular distribution. (b) Summed line density of the atoms trapped radially by the bottle beam. The step size of $\SI{174}{\micro\meter}$ along $z$ between recorded planes is indicated as the horizontal scale bar at lower right. The white vertical scale bar at left has a length of $\SI{50}{\micro\meter}$. (c) Number of radially trapped atoms in the bottle beam as a function of axial position. From this we estimate the length of the bottle as $l_{\rm{BB}}=\SI{1570\pm 174}{\micro\meter}$, taking the background of $\SI{500}{atoms}$ into account.}
\label{fig:atoms}
\end{figure}

For characterization of the effective potential along $z$ given by the bottle beam and the gravitational potential, a series of measurements are performed. 
The total magnification by which the bottle beam is reimaged into the vacuum chamber is needed for the calculation of the trapping frequencies and other trap parameters. 
Due to gravity, the axial equilibrium position of the atoms is dependent on the strength of the bottle-beam potential. 
Varying the optical power of the bottle beam shifts the equilibrium position in a known fashion, therefore leaving the magnification as a free parameter. 
To extract this information, a BEC is loaded into the bottle beam at maximal power. 
After an initial holding time, the optical power is lowered to the targeted value and an additional holding time is added to ensure equilibration. 
From the axial position of the atom cloud as a function of the optical power, the magnification is determined to $|M_{\rm tot}|=\SI{0.32\pm0.02}{}$ in agreement with the value of $|M^{\mathrm{th}}_{\rm tot}|=\SI {0.30\pm 0.02}{}$ calculated from the parameters of the optical reimaging system.\\
%
A good parameter to characterize the resulting potential is the axial trapping frequency in the harmonic approximation given by Eq.~\eqref{wz}.
The oscillation frequency for seven different starting positions is measured allowing for an additional quantification of the anharmonicity of the potential caused by gravity (see Fig.~\ref{fig:oscillations}). Starting with a BEC in the CDT, the power of the bottle beam is increased to the maximum value of $\SI{52}{mW}$ as described above, and the BEC is released into the bottle beam by suddenly switching off the CDT. 
By varying the displacement of the bottle beam relative to the CDT along the $z$ axis, the separation between the point of release and the equilibrium position can be modified. 
For shifting the bottle beam, the lens $\text{L}_3$ in Fig.~\ref{fig:setup} is moved along the beam axis, therefore shifting the image of the bottle beam. 
After release of the BEC into the bottle beam and a varying holding time, the atom distribution is detected by absorption imaging along the secondary detection path. 
A 2D-Gaussian fit is used to extract the position of the center of the atom cloud. 
In Fig.~\ref{fig:oscillations} the center positions along the $z$ axis for different times and initial axial displacements are shown. 
The data are fitted by a damped sinusoidal function to identify the frequency of the oscillation. 
We record oscillation frequencies of $\omega_{z, \text{max}}=2\pi\times\SI{55,7\pm 0.8}{\hertz}$ for the smallest and $\omega_{z,\text{min}}=2\pi\times\SI{51.7\pm 0.2}{\hertz}$ for the largest oscillation amplitude. 
The oscillation frequencies show a shift towards lower values for larger amplitudes. 
This can be explained by the increased anharmonicity of the potential for a larger deviation from the equilibrium position. 
Using Eq.~\eqref{wz},
we find $\omega_{z}^{\rm{th}}=2\pi\times\SI{57\pm 13}{\hertz}$. 
For the calculation, the measured values of the  beam waist, Rayleigh range, and magnification are used for imaging from the focal plane to the plane of the atoms. 
With Eq.~\eqref{wr} we calculate a radial trapping frequency of $\omega_{r}^{\rm{th}}=2\pi\times\SI{2881\pm 730}{\hertz}$ for the trap parameters used. 
Since the starting point of the oscillation is fixed in the absolute reference system by the position of the CDT and the bottle beam is shifted relative to it for the different measurements in Fig.~\ref{fig:oscillations}, all oscillations start at the same point. 
The position of the equilibrium point of the potential can be extracted from the offset of the sinusoidal fits. This can be used to validate the magnification of the telescope consisting of $\text{L}_4$ and $\text{L}_5$. 
We find a magnification of $|M_{4,5}|=\SI{0.59\pm 0.03}{}$ which confirms the theoretical value $|M_{4,5}^{\rm{th}}|=\SI{0.60 \pm 0.03}{}$.\\ 
%
%
%
%
In order to characterize the spatial structure of the bottle beam in the vacuum chamber, we tomographically record the atom distribution as a function of $z$. Planes of finite thickness are selected utilizing a light sheet (LS) potential in the plane of the CDT. 
The radial structure of the dark-focus potential is mapped by imaging the atom distribution in the plane selected by the LS.
The LS potential is created by a Gaussian beam which is elongated in one dimension. 
Using the red-detuned light of a tapered amplifier laser system at $\lambda_{\rm{LS}}=\SI{783,5}{nm}$, the LS provides an attractive potential with a vertical trapping frequency of $\omega_{z}^{\rm LS}=2\pi\times\SI{169\pm 1,5}{\hertz}$. 
Calculating the Thomas-Fermi radius as well as the width of a thermal cloud in this potential leads to a thickness of less than $d=\SI{10}{\micro\meter}$ in the LS potential.
The planes are scanned by varying the axial position of lens $\text{L}_3$ therefore shifting the image of the bottle beam relative to the LS. 
Moving the lens by steps of $\SI{500\pm10}{\micro\meter}$ results in a displacement of the bottle beam with step size $\Delta z=\SI{174\pm18}{\micro\meter}$ in the vacuum chamber. 
Compared to this step size, the atomic layer in the LS is thin thus sampling only a small axial part of the potential. 
Loading the BEC from the CDT into the LS and adding enough time for expansion results in an oblate atom distribution larger than the bottle beam.
After expansion, the bottle beam is turned on instantaneously with low optical power since the support against gravity is achieved by the LS. 
The blue-detuned bottle beam pushes the atoms into regions of low intensity, resulting in a distribution that shows the imprint of the bottle-beam potential. 
In Fig.~\ref{fig:atoms}(a), six examples of such distributions, recorded along the primary detection path, are shown for varying axial displacements of the bottle-beam focal plane relative to the LS. 
\begin{table}[t!]
\begin{tabular}{c|c|c} 
\hline\hline
 & Calculation & Light field \\ 
\hline
\rule{0pt}{10pt}
\multirow{5}{*}{\shortstack {Focal \\ plane}} & $w_0=\SI{44.0\pm 1.9}{\micro\meter}$ &  $w_0=\SI{43.6\pm 3.0}{\micro\meter}$\\
& $R_0=\SI{42.0\pm 1.9}{\micro\meter}$ & $R_0=\SI{41.2\pm 2.8}{\micro\meter}$\\
& $l_{\rm BB}^{\rm FP}=\SI{15.6\pm 0.7}{\milli\meter}$& $l_{\rm BB}^{\rm FP}=\SI{14.2\pm 0.5}{\milli\meter}$\\
 & $\omega_r^{\rm FP}=2\pi\times\SI{295\pm 65}{\hertz}$&$\omega_r^{\rm FP}=2\pi\times\SI{298\pm 10}{\hertz}$\\
& $\omega_z^{\rm FP}=2\pi\times\SI{1.87\pm 0.23}{\hertz}$&$\omega_z^{\rm FP}=2\pi\times\SI{1.89\pm 0.11}{\hertz}$\\
\hline
 & Calculation & Experiment \\ 
\hline
\rule{0pt}{10pt}
\multirow{7}{*}{\shortstack {Atom \\ plane}} & $w_0=\SI{14.0\pm 2.1}{\micro\meter}$& $ - $\\
& $R_0=\SI{13.4\pm 1.0}{\micro\meter}$ & $R_0=\SI{13.2\pm 1.3}{\micro\meter}$\\
& $l_{\rm BB}=\SI{1.60\pm 0.21}{\milli\meter}$& $l_{\rm BB}=\SI{1.57\pm 0.174}{\milli\meter}$\\
& $U_{\mathrm{trap}}$ = $k_{\mathrm{B}}\times\SI{166.5}{\micro\kelvin}$&$ - $\\
& $U_{\mathrm{trap},g}$ = $k_{\mathrm{B}}\times\SI{87.0}{\micro\kelvin}$&$ - $\\
& $\omega_r=2\pi\times\SI{2881\pm 730}{\hertz}$&$ - $\\
& $\omega_z=2\pi\times\SI{57\pm 13}{\hertz}$&$\omega_z=2\pi\times\SI{55.7\pm 0.5}{\hertz}$\\
\hline\hline
\end{tabular}
\caption{Comparison of measured and calculated parameters of the bottle beam: Experimental values are determined from the intensity distribution in the focal plane and the atom distribution in the atom plane. Calculated values in the focal plane: $l_{\rm BB}^{\rm FP}$, $\omega_r^{\rm FP}$, and $\omega_z^{\rm FP}$ are determined for $\rho_0=\SI{0.96\pm 0.08}{}$ using Eqs.~\eqref{wr} and \eqref{wz} and the experimental values for $w_0$ and $z_{\mathrm{R}}$ as input; we do not apply the experimental value of $R_0$ for calculations since for its determination an assumption for the value of $\rho_0$ has to be made, leading to large uncertainties. Calculated values in the atom plane: $w_0$, $R_0$, $l_{\rm BB}$, $U_{\mathrm{trap}}$, $U_{\mathrm{trap},g}$, $\omega_r$, and $\omega_z$ are based on the calculated values in the focal plane and the magnification $|M_{\rm tot}|=\SI{0.32\pm0.02}{}$. The potential depth $U_{\mathrm{trap},g}$ includes the effect of gravity. 
\label{Tab_1}}
\end{table}
The annular atom distributions that can be seen in Fig.~\ref{fig:atoms}(a) far away from the axis result from atoms not being trapped radially inside the bottle beam but pushed outside by the blue-detuned light. These atoms are ignored in further analysis. 
For small displacements [images (2)--(4)] the bottle beam exhibits the characteristic central dark region allowing for a significant central atom distribution to exist.
The atom number is extracted from the 2D density plot of each plane in Fig.~\ref{fig:atoms}(a) and plotted over the corresponding position 
in Fig.~\ref{fig:atoms}(c).
In Fig.~\ref{fig:atoms}(b), a density plot is shown for the one-dimensional (1D) line density along the $x$ axis of the central part of the atom distribution. 
From this, the radius of the bottle beam can be determined to $R_0=\SI{13.2\pm 2.6}{\micro\meter}$.
Even for large displacements along $z$ [images (1), (5), and (6)], a non vanishing central atom distribution can be found.  
This is due to atoms being pushed out of the LS along $z$, still remaining inside the radial bottle-beam potential, 
and thus being recorded in the absorption image which integrates over all atoms along the $z$ axis. 
This results in an offset of about $\SI{500}{atoms}$ outside the trapping region in Fig.~\ref{fig:atoms}(c). 
From the points where the atom number rises above this background level, the length of the bottle-beam potential at the atom plane is estimated to $l_{\rm{BB}}=\SI{1.570\pm 0.174}{\milli\meter}$, which is in good agreement with the value of $l_\text{BB}^\text{th}=\SI{1.60\pm 0.21}{\milli\meter}$ derived from the calculated value in the focal plane. 
In Table~\ref{Tab_1}, we summarize measured and calculated values of the bottle-beam parameters. All experimental parameters show good agreement with the calculated values for the intensity distribution in the focal plane as well as the bottle-beam potential in the atom plane.
\subsection{Lifetime of atoms in dark-focus trap}
\begin{figure}[ht!]
\centering
\includegraphics[width=1 \columnwidth]{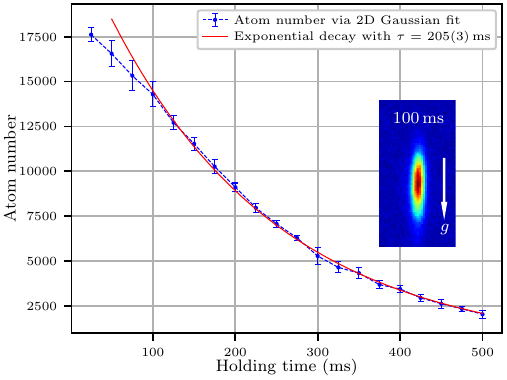}
\caption{Atom number as a function of holding time in the bottle-beam potential showing an exponential decay for trapping times larger than $\SI{100}{\milli\second}$. For trapping times smaller than $\SI{100}{\milli\second}$, a deviation from the exponential behavior is visible. Therefore, the exponential fit is carried out for $t\geq\SI{100}{\milli\second}$. From the fit (solid red line), a lifetime of $\tau=\SI{205\pm3}{\milli\second}$ is extracted. The inset shows the density plot of the atom distribution for $t=\SI{100}{\milli\second}$ after $t_{\rm TOF}=\SI{2}{\milli\second}$.}
\label{fig:lifetime}
\end{figure}

We measured the lifetime of the atoms in the dark-focus potential through absorption imaging for varying holding times up to $\SI{500}{\milli\second}$. 
To distinguish between atoms within and outside of the bottle beam, we utilize the secondary detection direction. 
A time of flight of $t_{\rm TOF}=\SI{2}{ms}$ prior to imaging decreases the optical density of the atom cloud and therefore reduces unwanted full absorption of the imaging light.
Through a 2D Gaussian fit to the density distributions, the number of atoms remaining in the potential is determined.
The resulting atom numbers are shown in Fig.~\ref{fig:lifetime} for different holding times.
An exponential function $N(t)=N_0\exp\left(-t/\tau\right)$ (solid red line) is fitted to the data. 
For times smaller than $\SI{100}{ms}$ the data deviate from a pure exponential behavior and therefore are not included in the fitting procedure. 
A possible reason for this deviation is the high density of the atom ensemble for short times in the bottle beam. 
For densities high enough to result in almost full absorption of the detection light, additional atoms cause sub proportional extra absorption, 
therefor leading to an underestimation of the total atom number.
From the exponential fit, we extract a lifetime of $\tau=\SI{205\pm3}{ms}$, which is primarily limited by gravitational loss and photon scattering. 
The equilibrium point of the potential is shifted away from the focal plane due to gravity. 
This leads to a higher intensity seen by the atoms and an increased scattering rate. 
Since the bottle beam is directed downwards parallel to gravity, the momentum transferred to the atoms propagates them further downwards along the direction of gravity out of the potential minimum and increases the scattering rate even more. 
This continues until the atoms are lost completely from the potential.
Orienting the bottle beam anti parallel to gravity would reduce this problem. 
Ultimately one would orient the bottle beam perpendicular to gravity.
\section{Evaluating different trap configurations}
\label{sec:evaluation}
The trap geometry realized and characterized in the experimental part of the previous sections is presenting the worst possible mode of implementation: The weak axial direction of the potential is oriented along the direction of gravity, the gravitational sag pulls the atoms out of the intensity minimum, and radiation pressure from spontaneous scattering from the downwards directed bottle beam accelerates the atoms along the direction of gravity in addition. Nevertheless, trapping of a BEC in this configuration has been achieved as presented in Sec.~\ref{sec5}.\\
In the current section, we discuss a series of standard configurations used in cold-atoms experiments and perform a comparison between the dark-focus trap and a corresponding focused Gaussian beam trap. We assume that the beam axis is oriented perpendicular to gravity and gravitational sag does not need to be considered. This requirement also holds for the Gaussian beam trap in which the axial curvature is even weaker, as shown below.
For any given Gaussian input beam of waist $w_0$, wavelength $\lambda$, and power $P$, the potential depth and the radial and axial trapping frequencies are given by
\begin{equation}
    U_\text{G} = \Tilde{U}_{\mathrm {0}}\frac{2P}{\pi w_0^2} = \Tilde{U}_{\mathrm {0}} I_0
\end{equation}
\begin{equation}
\omega_{r, \text{G}}=\sqrt{\frac{8\tilde{U}_{\mathrm 0}P}{\pi m w_0^4}}
\end{equation}
\begin{equation}
    \omega_{z, \text{G}}
= \sqrt{\frac{4\tilde{U}_{\mathrm 0} P{\lambda}^2}{\pi^3 m w_0^6}}.
\end{equation}
In Table~\ref{tab:comparison} we compare the Gaussian and the dark-focus ($\rho^{\rm{DF}}_0 = 0.924$) potentials. 
We have chosen experimentally accessible values for wavelength and optical power.
Three different configurations are presented: (a) traps for a BEC with a detuning of 1 nm and a laser power of 1 mW, (b) traps for a BEC in crossed-beam configurations with parameters as in the CDT used in our experiment for BEC production, and (c) optical tweezer traps as used to store individual atoms for quantum information processing.
To characterize both potentials in all three configurations, we calculate radial and axial trapping frequencies $\omega_{r}$ and $\omega_{z}$, the 3D geometric mean $\Bar{\omega}_{\mathrm{ho}}$, and the potential depth $U_{\mathrm{trap}}$. 
For comparability, we force that identical radial trapping frequencies are achieved in the Gaussian and in the bottle beam. 
With respect to a given input power $P_{\rm G}$ of the Gaussian beam at wavelength $\lambda_{\text{G}}$, one can calculate the necessary power in the bottle beam via
\begin{equation}\label{power_ratio}
P_{\rm BB}=\frac{1}{0.383^2} \left|\frac{\tilde{U}_{0\rm ,G}}{\tilde{U}_{0\rm ,BB}}\right| P_{\rm G} = 6.82 \left|\frac{\tilde{U}_{0\rm,G}}{\tilde{U}_{0\rm ,BB}}\right| P_{\rm G}.
\end{equation}
As can be seen in Table~\ref{tab:comparison}, this also leads to comparable trap depths. 
To demonstrate the advantage of the dark-focus trap, we evaluate the mean photon scattering rate $\Bar{\Gamma}$ for the resulting spatial atom distribution.\\
The two BEC configurations (a) and (b) are based on a BEC of $^{87}$Rb atoms in the Thomas-Fermi regime. For this approximation, an atom number of $N=\SI{20000}{}$ in the $F=1$ hyperfine state and the scattering length $a_{11}=\SI{100.4}{a_0}$ \cite{egorov:2013:scattering} with the Bohr radius $a_0$ are used. The beam waist is $w_0 = \SI{38.0}{\micro\meter}$.
In configuration (a), we choose a detuning $|\Delta\lambda_{\mathrm{BB}}| = |\Delta\lambda_{\mathrm{G}}| = \SI{1}{\nano\meter}$ from the D1 line of rubidium at $\lambda_{\rm D1}=\SI{794.98}{\nano\meter}$ and a power $P_{\mathrm{G}} = \SI{1}{\milli\watt}$. For identcial $\omega_r$, a power $P_{\mathrm{BB}} = \SI{9.00}{\milli\watt}$ is required according to Eq.~\eqref{power_ratio}.
The trap depths in both beams are very similar, but the bottle beam provides a larger axial trapping frequency $\omega_{z}$ and a mean photon scattering rate $\Bar{\Gamma}_{\mathrm{TF}}$ reduced by about a factor of $700$ when averaged over the 3D Thomas-Fermi distribution.\\
In configuration (b), we take the parameters of the final stage of evaporative cooling in our CDT for BEC production with a fiber laser at $\SI{1070}{\nano\meter}$ \cite{lauber:2011:pra} and keep the power fixed at $P_{\mathrm{G}} = P_{\mathrm{BB}} =\SI{37}{\milli\watt}$. For the bottle beam, we determine the wavelength at which the radial vibrational frequencies match to $\SI{758.5}{\nano\meter}$. This wavelength is easily accessible with a TiSa laser and allows to transfer the BEC adiabatically into a nearer-resonant blue-detuned light field for further applications such as dipole-potential-guided atom experiments in ring traps \cite{turpin:2015:} or atomtronics \cite{amico:2017:atomtronics,amico:2021:roadmap} configurations. Again, the mean photon scattering rate is reduced, in this case by a factor of $5$. This factor can easily be increased by allowing an increased power $P_{\mathrm{BB}}$ and accordingly increased detuning $\lambda_{\mathrm{BB}}$. Confining atoms in such a dark-focus CDT could allow for a higher efficiency in evaporative cooling and increased lifetime of the BEC.\\
\begin{table}[ht!]
    \centering
    \begin{tabular}{c|cc}
    \hline\hline
    Application & \multicolumn{2}{c}{Light-field parameters and calculated} \\
       case  & \multicolumn{2}{c}{values for configuration based on} \\
         & Gaussian beam & Bottle beam\\
    \hline
    (a) BEC in & $P_{\mathrm{G}} = \SI{1}{\milli\watt}$ & $P_{\mathrm{BB}} = \SI{9.00}{\milli\watt}$\\
      single-beam trap & $\lambda_\mathrm{G}=\SI{795.98}{\nano\meter}$ & $\lambda_\mathrm{{BB}}=\SI{793.98}{\nano\meter}$\\
    \hline
        $\omega_{r}$ & $2\pi\times\SI{176.3}{\hertz}$ & $2\pi\times\SI{176.3}{\hertz}$\\
        $\omega_{z}$ & $2\pi\times\SI{0.831}{\hertz}$ & $2\pi\times\SI{1.083}{\hertz}$\\
        $\Bar{\omega}_{\mathrm{ho}}$ & $2\pi\times\SI{29.6}{\hertz}$ & $2\pi\times\SI{32.3}{\hertz}$\\
        $U_{\mathrm{trap}}$  & $k_{\mathrm{B}}\times\SI{-4.63}{\micro\kelvin}$ & $k_{\mathrm{B}}\times\SI{4.20}{\micro\kelvin}$\\
        $\Bar{\Gamma}_{\mathrm{TF}}$ & \SI{6.55}{\frac{1}{\second}} & \SI{9.19e-3}{\frac{1}{\second}}\\
    & & \\
    (b) BEC in  & $P_{\mathrm{G}} = \SI{35}{\milli\watt}$ & $P_{\mathrm{BB}} = \SI{35}{\milli\watt}$\\
    crossed-beam trap & $\lambda_\mathrm{G}=\SI{1070}{\nano\meter}$ & $\lambda_\mathrm{{BB}}=\SI{758.5}{\nano\meter}$\\
    \hline
        $\omega_{r}$ (one beam) & $2\pi\times\SI{124.8}{\hertz}$ & $2\pi\times\SI{124.9}{\hertz}$\\
        $\omega_{z}$ (one beam) & $2\pi\times\SI{0.791}{\hertz}$ & $2\pi\times\SI{0.733}{\hertz}$\\
        $\Bar{\omega}_{\mathrm{ho}}$ & $2\pi\times\SI{140.0}{\hertz}$ & $2\pi\times\SI{140.2}{\hertz}$\\
        $U_{\mathrm{trap}}$ & $k_{\mathrm{B}}\times\SI{-2.32}{\micro\kelvin}$ &     $k_{\mathrm{B}}\times\SI{3.16}{\micro\kelvin}$\\
        $\Bar{\Gamma}_{\mathrm{TF}}$ & \SI{16.1e-3}{\frac{1}{\second}} & \SI{3.05e-3}{\frac{1}{\second}}\\
    & & \\
    (c) Single-atom & $P_{\mathrm{G}} = \SI{10}{\milli\watt}$ & $P_{\mathrm{BB}} = \SI{10}{\milli\watt}$\\
        tweezers & $\lambda_\mathrm{G}=\SI{1051}{\nano\meter}$ & $\lambda_\mathrm{{BB}}=\SI{759}{\nano\meter}$\\ 
    \hline
        $\omega_{r}$ & $2\pi\times\SI{98479}{\hertz}$ & $2\pi\times\SI{98477}{\hertz}$\\
        $\omega_{z}$ & $2\pi\times\SI{23285}{\hertz}$ & $2\pi\times\SI{21976}{\hertz}$\\
        $\Bar{\omega}_{\mathrm{ho}}$ & $2\pi\times\SI{60896}{\hertz}$ & $2\pi\times\SI{59732}{\hertz}$\\
                $U_{\mathrm{trap}}$ & $k_{\mathrm{B}}\times\SI{-1001}{\micro\kelvin}$ & $k_{\mathrm{B}}\times\SI{907}{\micro\kelvin}$\\
        $\Bar{\Gamma}_{\mathrm{GS}}$ & \SI{3.87}{\frac{1}{\second}} & \SI{11.7e-6}{\frac{1}{\second}}\\
    \hline\hline
    \end{tabular}
    \caption{Comparison of Gaussian and bottle-beam based traps for BECs (a,b) and single atoms (c). In all configurations, the dark-focus trap exhibits reduced rates for spontaneous scattering at comparable trapping properties. (See text for details.)}
    \label{tab:comparison}
\end{table}
In configuration (c), we compare optical tweezers for single atoms as used in experiments for quantum simulation, computation, and metrology. Again, we apply typical experimental parameters, i.e., beam waist $w_0 = \SI{1}{\micro\meter}$ and trap depth $|U_{\mathrm{trap}}| = k_{\mathrm{B}}\times\SI{1}{\milli\kelvin}$ (see, e.g., Refs. \cite{ohldemello:100atoms,schlosser:2023:multilayer}). We assume that an array of $100$ traps shall be created out of an available laser power of $\SI{1}{\watt}$, giving $P_{\mathrm{G}} =  P_{\mathrm{BB}} = \SI{10}{\milli\watt}$. The atoms shall be cooled to the 3D vibrational ground state, which allows to calculate the mean scattering rate $\Bar{\Gamma}_{\mathrm{GS}}$ using the ground state wave function as atom distribution. Since the single atom is localized at the potential minimum to a very high degree, spontaneous scattering in the blue-detuned trap is reduced by a factor of $\SI{3e5}{}$. This presents a tremendous advantage concerning the required coherence times for quantum technology applications. In addition, it has been shown that atoms in a highly excited Rydberg state such as those used for quantum gate operations can be trapped in blue-detuned bottle-beam traps \cite{barredo:2020:rydberg}. 
This makes the CR-based dark focus an advantageous configuration of individual-atom tweezer experiments.\\
%
%
\section{conclusions}
\label{conclusions}
We have demonstrated the experimental implementation of a blue-detuned 3D atom trap obtained from a single focused Gaussian beam through the CR phenomenon in biaxial crystals. We have deduced simple formulas for the trapping frequencies and potential barriers in three dimensions as a function of typical experimental parameters.
One of the advantages of this configuration is that CR provides the full conversion of the input power into the 3D dark-focus beam and avoids conversion losses, in contrast to other methods, e.g., based on SLMs, which introduce losses due to diffraction in the generation of the underlying Laguerre-Gaussian beams. Moreover, biaxial crystals can be transparent in an extremely wide spectral range \cite{darcy:2013:oe} [e.g., wavelengths of $\SI{350}{\nano\meter}$ --$\SI{5,5}{\micro\meter}$ in KGd(WO$_4$)$_2$], in contrast to SLMs, which only work in a narrow spectral range, usually a few hundreds of $\rm{nm}$. These features make the 3D dark-focus beam, produced by CR, a very useful tool for particle manipulation \cite{turpin_vault:2013:oe,krolikowski:2014:np} and atom trapping \cite{turpin:2015:}.\\
Further applications can be expected: If instead of a Gaussian symmetric input beam, an elliptical beam is used, the 3D dark focus will lead to a pair of elliptical beams divided by a thin dark region. This configuration could be used as a dark light sheet potential, where cold atoms are trapped in between the bright regions. The combination of the 3D dark focus with an array of microlenses \cite{dumke:2002,ohldemello:100atoms} would lead to the generation of a 2D array of 3D dark-focus traps being of significant interest for atom trapping in quantum computing and simulation experiments. Finally, changing the control parameter to $\rho_0 \gtrsim 1$, the dark focus evolves into a dark ring \cite{turpin:2015:}, which might be even interfered with a plane wave to give a 1D stack of doughnut-like dark-minimum potentials, ideal for quantum many-body experiments \cite{PhysRevA.88.063627,PhysRevA.102.023331,bazhan:2022}.\\
%
\begin{acknowledgements}
The authors gratefully acknowledge financial support through MCIN/AEI/10.13039/501100011033 Grant No. PID2020-118153GB-I00, the Catalan Government (Contract No. SGR2021-00138), and DAAD Contracts No.50024895 and No. 57059126. A.T. acknowledges financial support through Grant No. AP2010-2310 from the MICINN and DAAD Grant No. 91526836. 

\end{acknowledgements}

\bibliography{Bottle_Beam.bib}

\end{document}